\documentclass[%
 aip,
 amsmath,amssymb,
reprint, twocolumn 
]{revtex4-1}

\usepackage{graphicx}
\usepackage{dcolumn}
\usepackage{bm}

\usepackage[utf8]{inputenc}
\usepackage[T1]{fontenc}
\usepackage{mathptmx}
\usepackage{etoolbox}
\usepackage{xcolor}
\usepackage{soul}

\makeatletter
\def\@email#1#2{%
 \endgroup
 \patchcmd{\titleblock@produce}
  {\frontmatter@RRAPformat}
  {\frontmatter@RRAPformat{\produce@RRAP{*#1\href{mailto:#2}{#2}}}\frontmatter@RRAPformat}
  {}{}
}%
\makeatother
\begin{document}

\preprint{AIP/123-QED}

\title{Predictive first-principles simulations for co-designing next-generation energy-efficient AI systems}
\author{Denis Mamaluy}
 \affiliation{Sandia National Laboratories, Albuquerque, NM 87123, USA}
 \email{mamaluy@sandia.gov}
\author{Md Rahatul Islam Udoy}%
 \affiliation{Department of EECS, University of Tennessee, Knoxville, TN 37996, USA}
 
\author{Juan P. Mendez}%
  \affiliation{Sandia National Laboratories, Albuquerque, NM 87123, USA}
\author{Ben Feinberg}%
  \affiliation{Sandia National Laboratories, Albuquerque, NM 87123, USA}
\author{Wei Pan}%
  \affiliation{Sandia National Laboratories, Livermore, CA 94551, United States}
\author{Ahmedullah Aziz}%
 \affiliation{Department of EECS, University of Tennessee, Knoxville, TN 37996, USA}
   

\date{\today}

\begin{abstract}
In modern generative-AI workloads, matrix-vector/matrix-matrix multiplications (\emph{MatMul}) dominate the compute and energy cost. Achieving dramatic reductions in energy per token therefore requires a novel, specialized hardware that is co-designed across materials, devices, interconnects, circuits, and architectures rather than optimized at any single layer in isolation. In this \emph{Perspectives} article, we argue that \emph{predictive} (first-principles, fitting-parameter-free) device and interconnect simulations can close the loop between nanoscale physics and workload-level metrics, enabling the identification of device/interconnect operating regimes that plausibly support \emph{orders-of-magnitude} improvements in energy efficiency of AI accelerators.
\end{abstract}

\maketitle

\section{Introduction}
The sharp increase of energy consumption in AI applications is a growing concern, as highlighted by the International Energy Agency (IEA) \cite{IEA2025_EnergyAI} and the U.S. Department of Energy\cite{DOE_AI_DataCenter_2024}. These reports indicate that the energy use by data centers is projected to increase unsustainably, driven largely by the growing demands of AI. Worldwide efforts are focused on mitigation measures to reduce environmental impact. For instance, the DOE’s AMMTO recently published a roadmap targeting a 100-fold improvement in microelectronics energy efficiency over the next decade and 1000x over the next two decades\cite{DOE_EES2_2024}. An eco-friendly alternative to the brute-force ramping up of energy production, which may be a highly non-trivial task in post-industrial societies\cite{Ripa2021_EnergyMetabolismPostIndustrial},
is to increase the energy efficiency of the AI-related computations themselves. More specifically, it is highly desirable to reduce the energy spent on average per operation in AI-related tasks, assuming that the total computational time remains at least unchanged (or is reduced). Additionally, this approach would help mitigate another common challenge faced in data centers, namely heat dissipation\cite{ThermalPerformanceDataCenters}.

\begin{figure}[h!]
    \centering
    \includegraphics[width=1\linewidth]{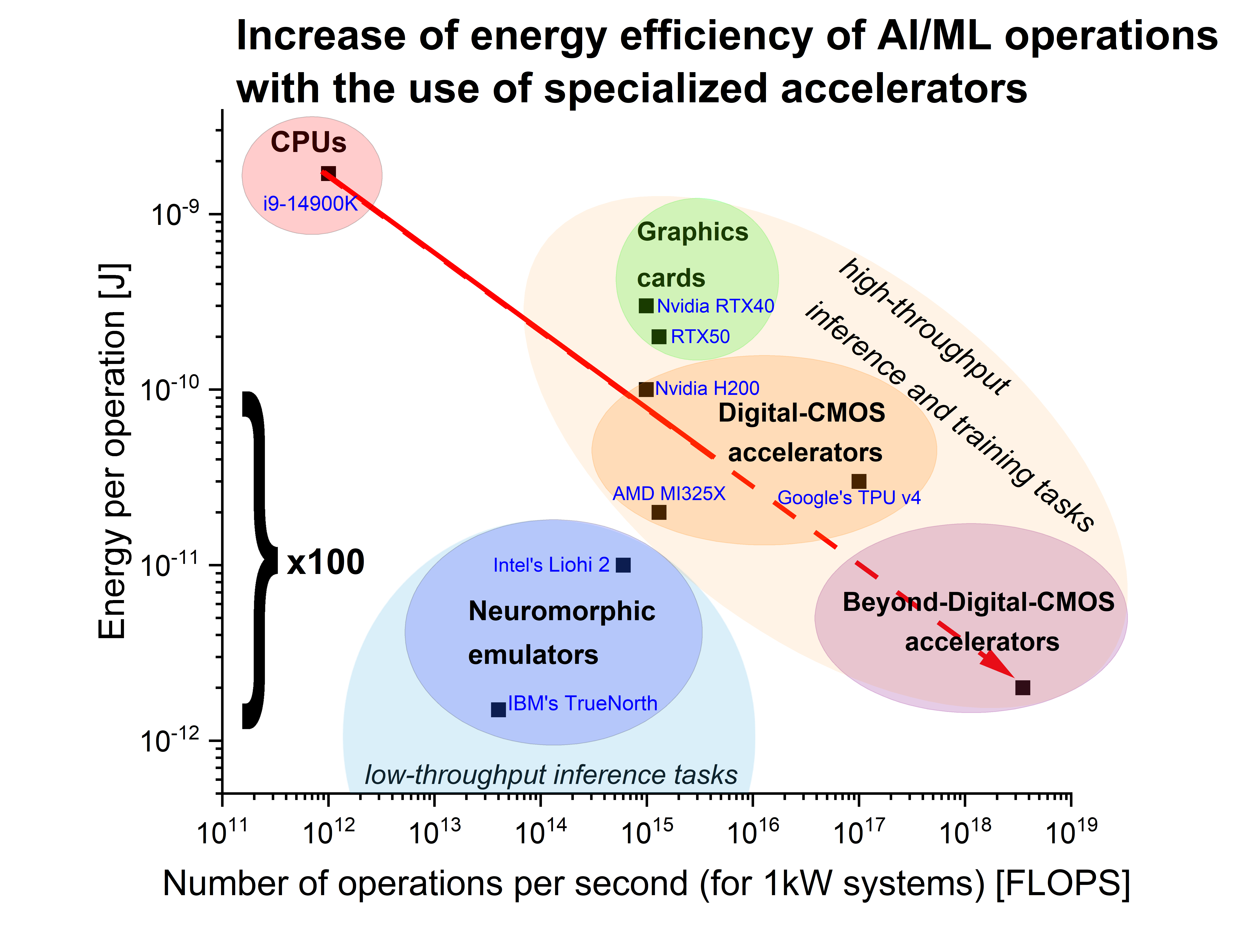}
    \caption{Relation between the energy spent per operation and the number of operations per second across different computing systems (assuming each consumes 1kW of power). AI-related applications of the future will demand even higher throughput and therefore must be based on much more energy-efficient computing systems (denoted as "Beyond-Digital-CMOS accelerators"). The energy scale of this region, roughly $1$--$100$~fJ per operation, is motivated quantitatively in Figure~\ref{fig:energy_ladder}.}
    \label{fig:Energy vs computation}
\end{figure}

The craving for the increased computational power in consumer applications and the corresponding need to increase their energy efficiency has already led to the spectacular rise of "GPU computing".
Graphics Processing Units (GPUs) are designed to deliver high‑throughput execution of a narrow set of linear‑algebra operations, achieving far greater energy efficiency than scaling out the same operations on CPUs.
Consequently, for certain highly parallel tasks (such as those in AI applications), the increased energy efficiency of GPU-computations also greatly increases the number of operations per second compared to CPUs as illustrated in Figure~\ref{fig:Energy vs computation}. 
Recently, in addition to "traditional" graphics cards, other specialized accelerators such as Tensor Processing Units (TPU)\cite{TPUv4} and Neural Processing Units (NPU)\cite{AMDmi325x} have emerged, indicating that the race for greater computational power, while achieving higher energy efficiency, has only begun. 

The energy consumption of AI-related computations consists of two categories: \emph{training} and \emph{inference} (execution of a trained model to generate outputs) \cite{AIbook} tasks. Although the energy consumption of inference applications grows over time and can eventually exceed that of training \cite{BLOOM_power_consumption, AI_power_consumption}, in practice, training accounts for, and will likely continue to account for, a very large portion of total energy use, driven by the ongoing competitive push to update the models. In this regard, it is important to note that while existing neuromorphic hardware emulators, some of which are shown in Figure~\ref{fig:Energy vs computation}, may be suitable for low energy real-time inference tasks, their relatively low throughput remains the issue \cite{2021LoihiSurvey}. Since these neuromorphic platforms are themselves implemented in digital CMOS, they cannot exceed CMOS limits at the device-physics level; any practical energy advantages typically come from architecture/algorithm effects (e.g., event-driven sparsity, reduced data movement, and lower activity factors) and are therefore workload-dependent. For large-model training and high-throughput dense linear-algebra workloads, throughput and software ecosystem constraints still favor GPU/TPU-class accelerators.
Thus, it can be argued that \emph{a new type} of computing hardware with the focus on energy efficiency is highly desired for both high-throughput inference and all AI training tasks. 
We can notice that this novel energy efficient hardware does not need to be restricted to neuromorphic approaches that focus on replicating \emph{actual} biological neural processes. Be it neuromorphic, analog or hybrid architecture - as long as it can provide sufficiently high throughput  and low energy cost per operation - it should be utilized. We denote this desired new hardware type as "Beyond-Digital-CMOS" accelerators in Figure~\ref{fig:Energy vs computation}. 

In this \emph{Perspectives} article, we envision the development of AI accelerators based on Beyond-Digital-CMOS devices that are co-designed to be extremely energy efficient in the specialized, highly scalable circuits. 




\section{Computational costs \& energy efficiency of GPT-like architectures}\label{sec:costs}

\begin{figure}[h!]
    \centering
    \includegraphics[width=1.1\linewidth]{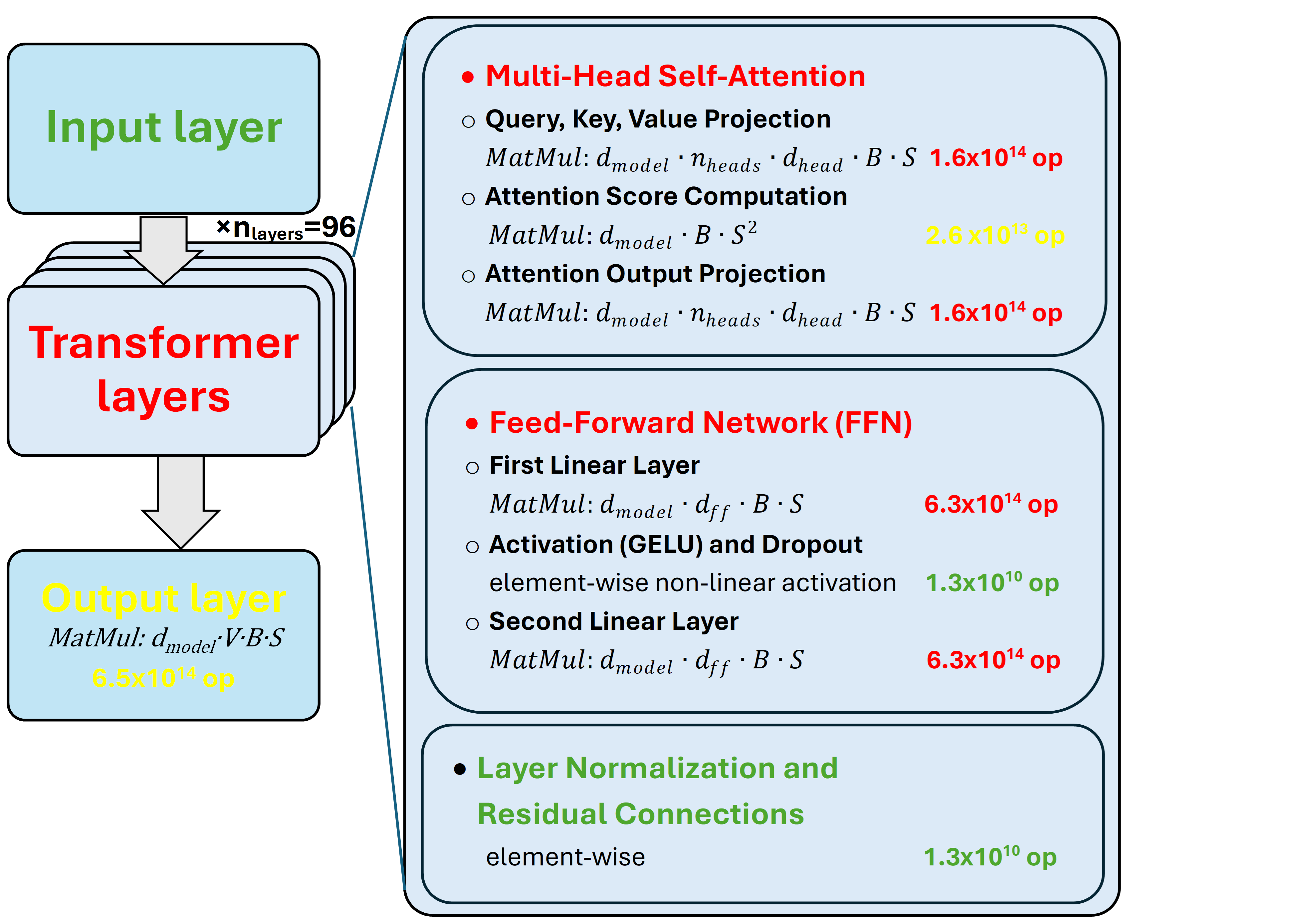}
    \caption{
    Schematics of GPT architecture and assessment of highest computational costs within it, shown as the approximate number of field operations (op) in the batched-training (GEMM) regime. The GPT (decoder-only Transformer) architecture\cite{GPT,Multi-head-Attention,FNN} is a stack of $n_{layers}$ identical layers, each comprising multiple components with varying computational costs. The most computationally expensive operations are color-coded by cost: red for high, yellow for medium, green for low. Related variables and their values in GPT-3 model\cite{GPT3} are: $B=512$, batch size; $S=2048$, sequence length; $d_{model}=12288$, the size of model embeddings and hidden states; $n_{heads} =96$, number of attention heads; $d_{head} =128$, dimension per attention head; $V=50257$, vocabulary size; $d_{ff}=49152$, feed-forward dimension (the hidden layer size in the feed-forward network, typically $4\times d_{model}$), $n_{layers}=96$, the number of stacked Transformer decoder blocks. 
    }
    \label{fig:GPT}
\end{figure}

Let us now discuss what constitutes the most energy-demanding calculations in the most successful framework for the generative AI, Generative Pre-trained Transformers (GPT)\cite{GPT}. In GPT-like frameworks, matrix-vector and matrix-matrix multiplications (\emph{MatMul}) dominate the computational workload. The operation counts in Figure~\ref{fig:GPT} correspond to batched training, where these products are executed as dense matrix-matrix multiplications (GEMM); during autoregressive inference the same attention and feed-forward layers reduce, per generated token, to matrix-vector products (GEMV) of identical structure and lower arithmetic intensity. Both regimes are therefore covered by the term \emph{MatMul} used throughout this article. As illustrated in Figure~\ref{fig:GPT}, the computational costs of \emph{MatMul} operations, especially in the multi-head self-attention, including QKV (Query, Key, Value) projection and attention output projection layers, as well as the feed-forward neural network (FNN)  and output layer, are orders of magnitude higher than other costs. Specifically, in self-attention\cite{Multi-head-Attention} and FNN \cite{FNN}, the \emph{MatMul} complexity scales as $O(B \cdot S \cdot d_{model}^2)$ and $O(B \cdot S \cdot d_{model} \cdot d_{ff})$, 
respectively, where $B$ is the batch size, $S$ is the sequence length, $d_{model}$ is the embedding size, and $d_{ff}$ is the feed-forward network dimension in the hidden layers. By contrast, the computational cost in other operations, such as \emph{Gaussian-Error-Linear-Unit (GELU) Activation}, normalization and residual connections scale as $O(B \cdot S \cdot d_{model})$. Additionally, experimental floating-point operation (FLOP) accounting and performance models for GPT-style Transformers show that dense \emph{MatMul} operations constitute the dominant fraction of total computation, while the relative importance of non-\emph{MatMul} operations can increase for memory-bound implementations or specific model/configuration regimes\cite{MegatronLM2021,PaLM2022}.

Thus optimization of the energy efficiency of Matrix-vector multiplications is the most crucial step for the overall increase of the energy efficiency and throughput of GPT-like architectures (Fig.~\ref{fig:Energy vs computation}). The creation of a specialized "MatMul-accelerator" would dramatically decrease the energy spent per token, possibly providing the necessary 100-fold gain in energy efficiency \cite{DOE_EES2_2024}. In practice, the MatMul energy and speed are set jointly by (i) \emph{dynamic} switching energy and drive capabilities of the active devices (e.g., $CV^{2}$, on-current, achievable operating frequency), (ii) \emph{static} energy losses such as leakage currents that accrue over the computation time, and (iii) data-movement and interconnect costs (wire $RC$ [delay], fan-out, and memory/interface energy). Therefore, device-only claims should be evaluated together with interconnect-aware compact models and workload-level metrics. 

\subsection*{How to build a more energy-efficient AI accelerator?} 
The digital CMOS technology continues to provide smaller device dimensions and higher device densities, and this process itself continues to enable the lower energy cost per logical operation (due to unchanged/lower computation time). However, the appetite for AI-related applications seems to grow at a much higher rate than the computing power that modern digital CMOS technology can provide\cite{IEA2025_EnergyAI}. Additionally, in modern digital CMOS computing systems energy dissipation happens primarily on the device level with  estimates weighing roughly up to $50\%$ each of energy dissipation to the devices and interconnects\cite{Interconnects_power}. Thus, optimizations on the device and material levels are crucial to enable energy-efficient, massively parallel operations. But how to design devices for the increased energy efficiency of such operations? The 63 year long\cite{WanlassSah1963} development of digital CMOS technology has led to creation of the specific transistor design rules that balance the need to keep the off-currents low, while enabling relatively high on-currents for the efficient "fan-out", i.e. powering-up the sufficient number of switches (other transistors). The key efficiency metrics of a transistor in a digital computation is its switching energy, $CV^{2}$, which, despite the effective cessation of the voltage $V$ scaling,  is still reduced with each new digital CMOS node as the gate capacitance $C$ is scaled down. Consequently, a good digital transistor must have a high on-to-off ratio, while providing sufficiently high drive current. However, for the alternative, neuromorphic\cite{Mead1990Neuromorphic}, analog\cite{Shannon1941Analog}, wave\cite{Wave-parallel-computing}, etc. computational approaches the requirements for a “good transistor” significantly change, with the increased emphasis for instance on the signal amplification and/or the quality of the linear regimes\cite{Wang2021}. Therefore, the energy consumption of an optimized alternative device can also be reduced compared to CMOS transistors optimized for digital computations. 
Overall, for alternative computational paradigms such as the analog '2T1C' scheme proposed in Ref.\cite{Wang2021}, the greatest opportunity for improving energy efficiency arises not from reducing the energy consumed by an individual transistor, but from dramatically reducing the number of transistors required to perform matrix-vector multiplications through elementary, highly scalable multiply-accumulate (MAC) operations.

\begin{figure}[h!]
    \centering
    \includegraphics[width=\linewidth]{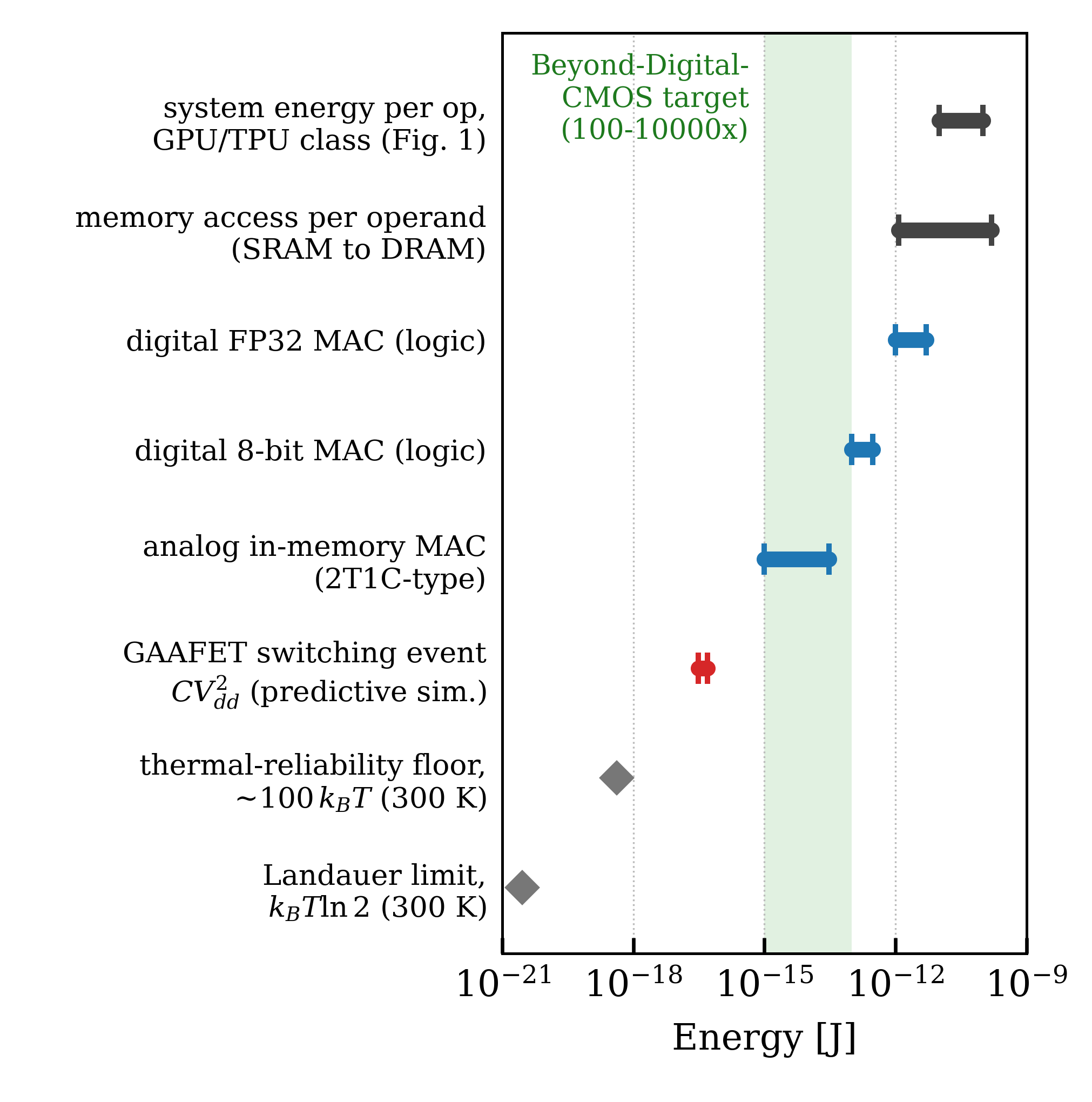}
    \caption{Energy scales relevant to MatMul acceleration, from the thermodynamic floor to the system level. For deterministic logic, the practical floor is set not by the Landauer bound but by the thermal-reliability requirement of ${\sim}100\,k_{B}T$ per switching event\cite{FET_downscaling_limit}. The switching energy of a single state-of-the-art GAAFET, $CV_{dd}^{2}\approx 30$--$50$~aJ, follows directly from the predictive first-principles simulations of Ref.~\cite{GAAFET_TechRxiv}: the gate capacitance $C\approx 64$~aF is computed from the simulated gate stack, and the upper edge of the bar accounts for parasitic overhead. A digital MAC operation nevertheless costs $0.1$--$5$~pJ in logic alone\cite{Horowitz2014}, and system-level energies per operation in GPU/TPU-class accelerators (cf.\ Figure~\ref{fig:Energy vs computation}) are further burdened by data movement\cite{Horowitz2014} and by facility-level overheads such as cooling and power delivery\cite{ThermalPerformanceDataCenters,IEA2025_EnergyAI}. The shaded band marks the Beyond-Digital-CMOS target of $1$--$100$~fJ per operation, $10^{2}$--$10^{4}$ times below today's system level\cite{DOE_EES2_2024}. Alternative MAC schemes such as the analog 2T1C approach\cite{Wang2021} fall within this band, but realizing them requires devices re-optimized for figures of merit beyond the digital on/off paradigm.}
    \label{fig:energy_ladder}
\end{figure}

To make the required gains quantitative, Figure~\ref{fig:energy_ladder} places the relevant energy scales on a single axis. The switching energy of a single state-of-the-art GAAFET is only $30$--$50$~aJ, obtained directly, without fitting parameters, from the predictive simulations discussed in Sec.~\ref{sec:predictive}. A complete digital MAC operation nevertheless requires $0.1$--$5$~pJ in logic alone\cite{Horowitz2014}, and the system-level energy per operation of GPU/TPU-class accelerators is $10$--$100$~pJ (Figure~\ref{fig:Energy vs computation}), with memory access and data movement contributing up to hundreds of picojoules per operand\cite{Horowitz2014}, and facility-level overheads (cooling, power conversion and delivery, idle units) multiplying the total by a further factor of roughly $1.1$--$1.5$\cite{ThermalPerformanceDataCenters,IEA2025_EnergyAI}. A complete accounting must therefore be performed at the wall plug, which is the normalization already used in Figure~\ref{fig:Energy vs computation}. The Beyond-Digital-CMOS region of Figure~\ref{fig:Energy vs computation} thereby acquires a physical definition: roughly $1$--$100$~fJ per operation, i.e., $10^{2}$--$10^{4}$ times below today's system level, a range that brackets the DOE AMMTO roadmap targets\cite{DOE_EES2_2024}. For deterministic logic, the operative lower bound is not the Landauer limit but the thermal-reliability floor of ${\sim}100\,k_{B}T\approx 0.4$~aJ per switching event\cite{FET_downscaling_limit}; statistically error-tolerant workloads, such as neural-network inference, can operate below it, which is part of the Beyond-Digital-CMOS opportunity. Two conclusions follow. First, this target cannot be reached by improving the digital switch alone: even the leanest 8-bit digital MAC merely touches the upper edge of the band, because a digital MAC comprises thousands of switching events together with the associated wiring. Second, MAC schemes that replace these thousands of events by a handful of charge transfers, such as the analog 2T1C approach\cite{Wang2021}, fall within the band at the device level; their realization, however, requires transistors and interconnects re-optimized for different figures of merit (linearity, retention, signal amplification) under leakage and variability constraints. Establishing these constraints quantitatively, without fitting parameters, is the role of predictive simulation, to which the remainder of this article is devoted.

\subsection*{Co-designing energy efficient hardware using predictive device simulations}

To achieve higher energy efficiency and performance, alternative architectures must be implemented using \emph{novel devices} optimized for energy efficient utilization in these non-digital or hybrid computations. A significant challenge presents the above-mentioned fact that ideal (desired) device characteristics of alternative, non-digital computing schemes are generally not known. This necessitates the process of \emph{co-design}, the term that is well known to computer architects for the microarchitecture-algorithm-application optimizations\cite{Co-design1,Co-design2}. Co-design optimizations can also be applied at the material-device-circuit levels, to provide unprecedented gains in speed and energy efficiency on the system level. 

The process of co-design makes it possible to determine the optimal device characteristics for the specific computational approach using feedback/forward optimization loops. The resulting optimal device characteristics can then be employed to address an \emph{inverse problem}, in which predictive device simulations are used to determine the optimal designs for the device, material, and doping profiles. More concretely, let $F:\mathbf{p}\mapsto\mathbf{c}$ denote the forward map realized by the predictive simulator, from design parameters $\mathbf{p}$ (geometry, material stack, doping profiles) to circuit-usable characteristics $\mathbf{c}$ (e.g., $I$-$V$ curves, capacitances, leakage components). The circuit- and system-level stages of the co-design loop supply target characteristics $\mathbf{c}^{*}$ with tolerances: for a digital cell, a required on-current at a given $CV^{2}$ and a leakage ceiling; for an analog MAC cell, additionally the linearity and dynamic range of the transfer characteristic. The inverse problem is then $\mathbf{p}^{*}=\arg\min_{\mathbf{p}\in P}\lVert F(\mathbf{p})-\mathbf{c}^{*}\rVert_{w}$, where the admissible set $P$ encodes fabrication constraints and the weighted norm encodes the relative importance of each characteristic. Since every evaluation of $F$ is a full first-principles simulation, the minimization is performed in practice over fast surrogates of $F$, such as the machine-learning compact models discussed in Sec.~\ref{sec:vision}, trained on predictive simulation data, with candidate optima verified by direct simulation. Such first-principles, fitting parameter-free simulators, can provide accurate electrical characteristics of both the state-of-the-art (or future) digital CMOS transistors and the alternative beyond-CMOS devices, which then can be used in a sophisticated simulation framework to assess the whole computing system performance (including its energy efficiency). In the following, we first describe examples of such predictive device simulations, followed by a vision for the co-design framework utilizing the benefits of such first-principles simulations. 

\section{Predictive first-principles simulations}\label{sec:predictive}

\begin{figure*}[!t]
    \centering
    \includegraphics[width=\textwidth]{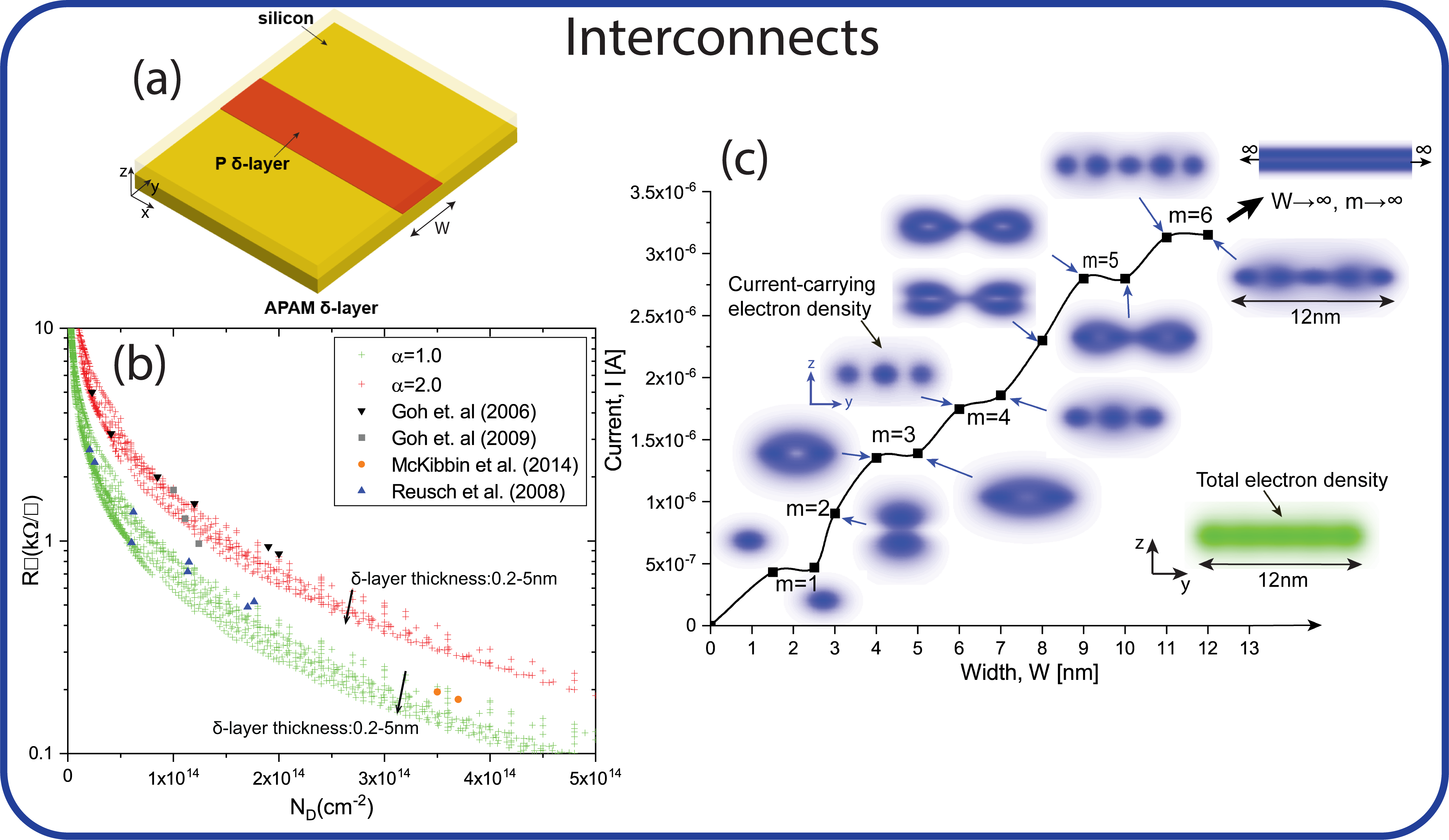}
    \caption{
    Predictive first-principles simulations of Si:P $\delta$-layer interconnects: (a) interconnect schematics; (b) predicted sheet resistances for different doping densities and thicknesses from Ref.~\cite{Mamaluy2021_CommPhys}, and comparison with measurements\cite{Goh:2006,Goh:2009,Reusch:2008,McKibbin:2014}; (c) predicted current (I) for different widths (W), and phosphorus sheet density of N$_{D}=10^{14}$cm$^{-2}$ from Ref.~\cite{DeltaLayer2022_SciRep}. The insets in blue color show the spatial distributions of current-carrying modes across a y-z plane, indicating the corresponding number of propagating modes (m). Inset in green color shows the total electron density all occupied electron states for a width W=12~nm.}

    \label{fig:Predictive_Simulaitons_Results_1}
\end{figure*}

\begin{figure*}[!t]
    \centering
    \includegraphics[width=\textwidth]{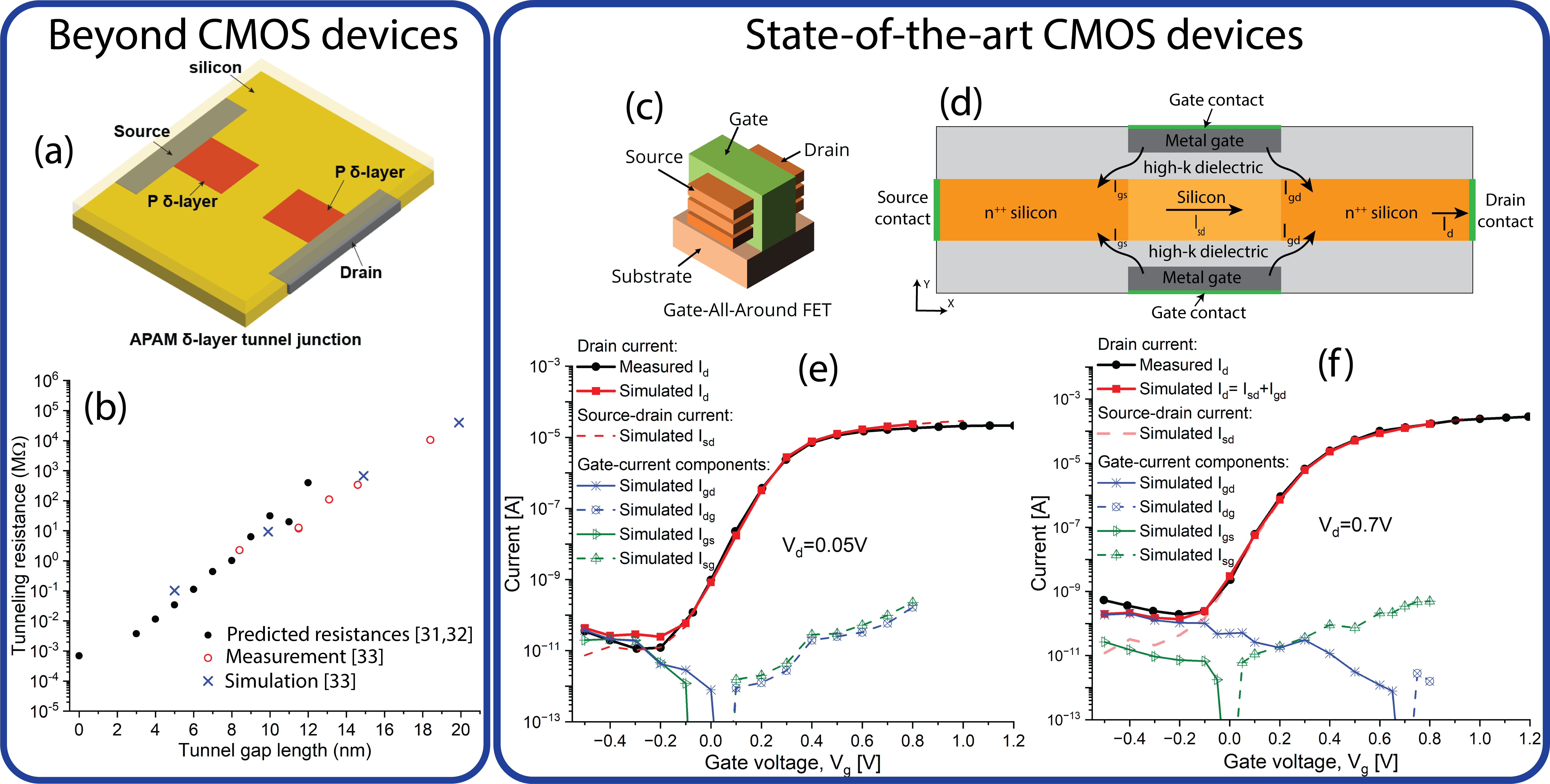}
    \caption{
    Predictive first-principles simulations for beyond-CMOS and state-of-the-art CMOS devices: (a) schematics of a $\delta$-layer tunnel junction device; (b) predicted tunneling resistances for a $\delta$-layer tunnel junction of thickness t=1~nm, width W=7~nm, and doping density of N$_D$=10$^{14}$cm$^{-2}$ from Refs.~\cite{DeltaLayer2022_SciRep,Mendez2023}, and compared against the measurement and simulations in Ref.~\cite{TJ_experiment};  (c)-(f) predictive simulations for conductive properties of GAAFET device\cite{GAAFET_TechRxiv}: (c) schematics of a three-nanosheets GAAFET; (d) schematics of a single nano-sheet channel simulated to investigate the leakage paths in GAAFETs shown in (e) and (f).}
    \label{fig:Predictive_Simulaitons_Results_2}
\end{figure*}

From the viewpoint of co-design, the role of predictive first-principles simulation is not merely to reproduce known or predict accurately new device behavior, but to provide a trustworthy mapping from \emph{design knobs} (geometry, materials, interfaces, doping profiles, and interconnect layout) to \emph{circuit-usable metrics} (delay/energy, parasitics, variability, and operating envelopes). This mapping is critical because MatMul accelerators operate at the edge of multiple constraints simultaneously: dynamic switching energy and drive current set attainable throughput, static leakage and standby losses accumulate over the wall-clock time of training/inference, and interconnect parasitics and data movement often dominate system-level energy consumption. 

Throughout this article, we use \emph{predictive} simulation to mean a physics-based approach that (i) uses a parameter set fixed independently of the specific device being predicted (i.e., avoids device-by-device fitting), (ii) outputs experimentally relevant electrical characteristics (e.g., $I$-$V$ curves, capacitances, contact/interface resistances, variability trends) for a given geometry/material stack, and (iii) provides a clear validation/uncertainty strategy when compared to measurements\cite{MamaluyMicDAT2024}. 

Several established atomistic and first-principles transport frameworks address related problems and should be acknowledged here. Empirical tight-binding NEGF in the $sp^{3}d^{5}s^{*}$ basis, as implemented in the NEMO/OMEN family of codes\cite{Luisier2006,Steiger2011}, resolves individual atoms and treats band-structure effects with high fidelity; its orbital parameters, however, are calibrated to bulk band structures or to hybrid-functional reference calculations, so the approach is semi-empirical, even though the parameters are held fixed across devices. DFT-NEGF, as implemented in TranSIESTA\cite{Brandbyge2002,Papior2017} and QuantumATK\cite{Smidstrup2020}, is parameter-free in principle; in practice, local and semilocal functionals inherit the band-gap problem, corrections via hybrid functionals reintroduce an adjustable exchange fraction, and the computational cost of charge-self-consistent finite-bias solutions restricts applications to systems considerably smaller than experimentally relevant device cross-sections at realistic doping densities. Semiclassical (Boltzmann-type) conductivity models\cite{MayadasShatzkes1970} remain the workhorse for interconnect resistivity, but they do not apply where transport is controlled by quantization and tunneling (cf.\ Figure~\ref{fig:Predictive_Simulaitons_Results_1}c). The criterion we emphasize in this article is therefore not methodological exclusivity but the demonstrated validation record under definition (i)--(iii) above: quantitative agreement with measured transport characteristics at experimentally relevant device dimensions, doping levels, and biases, achieved without device-specific re-fitting, and including a prediction made before the corresponding experiment (see below). To our knowledge, comparable demonstrations have not been reported for the alternative frameworks. Conversely, nothing in the co-design framework proposed in Sec.~\ref{sec:vision} is specific to the CBR implementation: any simulator meeting criteria (i)--(iii) can serve as its first-principles stage.

Conductive properties of modern state-of-the-art electron devices and interconnects are dominated by quantum-mechanical effects\cite{Quantum_dominates_1,Quantum_dominates_2,Quantum_dominates_3}. 
The essence of a first-principles approach for these systems is the rigorous treatment of electron transport as an \emph{open-boundary} quantum problem\cite{Mamaluy2021_CommPhys}.
In particular, the current must be computed directly from the quantum-mechanical flux through open contacts, i.e., from the current-density operator $\mathbf{j}\propto \Psi\nabla\Psi^{*}-\Psi^{*}\nabla\Psi$, where $\Psi(\mathbf{r})$ denotes the single-particle states of the open system and $\Psi^{*}$ their complex conjugates. This flux vanishes identically for closed-system (bound) states and becomes ill-defined for extracting non-equilibrium device conductance from a purely periodic band-structure picture. 
This motivates an explicitly open-system, real-space, charge self-consistent Non-Equilibrium Green’s Function (NEGF) Keldysh formalism \cite{Keldysh:1965,Datta:1997}: the open-system Schr\"{o}dinger equation is solved together with electrostatics (Poisson), while electron-electron interactions are treated within the local density approximation (LDA), so that carrier density and potential are mutually consistent under bias and the current is obtained from flux at the contacts. Conceptually, such "charge self-consistent NEGF" approach can be viewed as a DFT-like framework generalized (using albeit a different variational principle) to open boundary conditions: the generally unknown kinetic energy functional, $T[n]$, is approximated, for instance, by an effective-mass tensor operator in free electron basis\cite{Mamaluy2021_CommPhys}, an "external potential" functional, $V[n]$, is perfectly defined by the device geometry, materials and doping profiles, and, finally, the electron-electron interaction term, $U[n]$, is given within the LDA, which provides sufficient fidelity for free electrons in mesoscopic systems\cite{Mendez2023,GAAFET_TechRxiv}. 
Where traditional approaches with closed/periodic boundary conditions failed to reproduce experimentally relevant conductive properties, the rigorous open-system charge self-consistent quantum transport treatment enabled predictive transport modeling for advanced devices and interconnects\cite{Mamaluy2021_CommPhys,DeltaLayer2022_SciRep,Mendez2023,PhysRevApplied_2023,ACS_2025, GAAFET_TechRxiv}.

Open-system quantum transport simulations, which are typically highly computationally demanding, can be carried out using the charge self-consistent NEGF that is implemented via the Contact Block Reduction (CBR) method\cite{Mamaluy:2003,Mamaluy_2004,Sabathil_2004,Mamaluy:2005,Khan:2007,Khan_2008,Gao:2014,Mendez:2021}. The CBR method allows for a very efficient calculation of the local density of states (that replaces "band-structure" for non-periodic systems), charge density, transmission function, currents of an arbitrarily shaped multi-terminal two- or three-dimensional open device and scales linearly $O(N)$ with the system size $N$.  Applying the CBR method to multi-terminal structures makes it possible to treat all current-carrying contacts (source, drain, and gate(s)) fully quantum mechanically, which in turn substantially improves the accuracy of leakage current predictions\cite{GAAFET_TechRxiv}.

Below we summarize three representative examples of predictive simulation that illustrate how first-principles physics can be propagated to experimentally testable electrical characteristics and, ultimately, to compact models suitable for circuit/system studies. Importantly, the need for first-principles open-boundary quantum transport treatment is not limited to interconnects; it is equally central to understanding tunnel junctions and strongly confined transistor geometries such as in GAAFETs. The examples serve a dual role. They are stringent tests of the predictive methodology, since transport in these systems is entirely quantum, and they are candidate primitives for ultra-low-energy compute fabrics: atomically thin, degenerately doped $\delta$-layer interconnects offer extreme wiring density at low resistivity, and $\delta$-layer tunnel junctions provide atomically abrupt nonlinear elements, while the GAAFET example establishes the digital CMOS baseline that any Beyond-Digital-CMOS approach must outperform. We do not claim that $\delta$-layer technology is the accelerator technology of choice; the claim of this article concerns the methodology by which any such candidate should be evaluated.

\subsection*{Revealing quantum effects in predictive nanoscale interconnect simulations}


As device dimensions shrink into the nanoscale regime, interconnects must scale proportionally, limiting both energy efficiency and performance. At these feature sizes, charge transport in interconnects is governed by quantum-mechanical effects, leading to size-dependent resistivity\cite{Huang:2008,Steinhogl:2002,Josell:2009,Hu:2023}, increased surface and grain-boundary scattering\cite{Huang:2008,Graham:2022,Dutta:2018,Hu:2023}, and pronounced interface effects that are negligible in bulk conductors but dominant at nanometer dimensions.

In Ref.~\cite{Mamaluy2021_CommPhys}, we demonstrated predictive modeling of nanoscale interconnect behavior with quantitative links to experimentally measurable transport characteristics, as shown in Figure~\ref{fig:Predictive_Simulaitons_Results_1}a-c, capturing the relevant quantum-mechanical effects and, in some cases, revealing them. For example, as Figure~\ref{fig:Predictive_Simulaitons_Results_1}c illustrates, the well-known conduction steps due to each new propagating mode for narrow-width $\delta$-layer interconnects are also accompanied by strongly quantized spatial distributions of current-carrying electrons\cite{DeltaLayer2022_SciRep}. In the presence of capping layers and related impurities, this effect can lead to higher scattering rates and reduced currents. Such predictive results provide the physically grounded interconnect parameters (e.g., effective resistivity and contact/interface contributions) needed for compact modeling and system-level energy-delay estimates.


\subsection*{Predicting the conductivity of beyond-CMOS devices (theory first, experiment later)}

A key benchmark for predictive capability is the ability to forecast a device property before the corresponding experiment is performed, then validate the prediction afterward. As illustrated in Figure~\ref{fig:Predictive_Simulaitons_Results_2}a-b, we demonstrated this "theory-first" predictive workflow for $\delta$-layer tunnel junctions in Refs. \cite{DeltaLayer2022_SciRep,Mendez2023}, where the conductivity was predicted and subsequently confirmed by experiment\cite{TJ_experiment}. Because transport in these structures is intrinsically quantum and can be dominated by tunneling and strong confinement, a predictive workflow must again rely on an open-boundary quantum-transport treatment rather than closed/periodic-system or semiclassical conductivity models.

Beyond the specific junction geometry and effects of defects\cite{PhysRevApplied_2023,Mendez2023}, the broader point is that the predictive framework outputs electrical characteristics directly relevant to circuit modeling (e.g., conductance in the relevant bias/temperature regimes) and can therefore guide selection of materials and interfaces for ultra-low-energy compute fabrics.

\subsection*{Predictive modeling of advanced CMOS transistors (GAAFETs)}

Finally, predictive simulation is equally important for state-of-the-art CMOS nodes, both to establish realistic baselines and to quantify where beyond-CMOS concepts must outperform. In recent work, we applied predictive modeling to gate-all-around field-effect transistors (GAAFETs) with the goal of investigating the origin of non-thermionic behavior in the deep sub-threshold regime \cite{GAAFET_TechRxiv}. As the very good match to experiment demonstrates (shown in Figure~\ref{fig:Predictive_Simulaitons_Results_2}c-f), strong electrostatic confinement, conduction band quantization, and new tunneling/leakage paths fully justify an open-boundary quantum-transport treatment as the natural "first-principles" baseline for predictive $I$-$V$ characteristics and leakage analysis.

In the context of co-design, these predictive transistor models serve as the calibrated "ground truth" for leakage-delay-energy tradeoffs and variability constraints that ultimately bound achievable system-level gains. 


\section{Our vision/A proposed mini-road map}\label{sec:vision}

\begin{figure*}[!t]
    \centering
    \includegraphics[width=\textwidth]{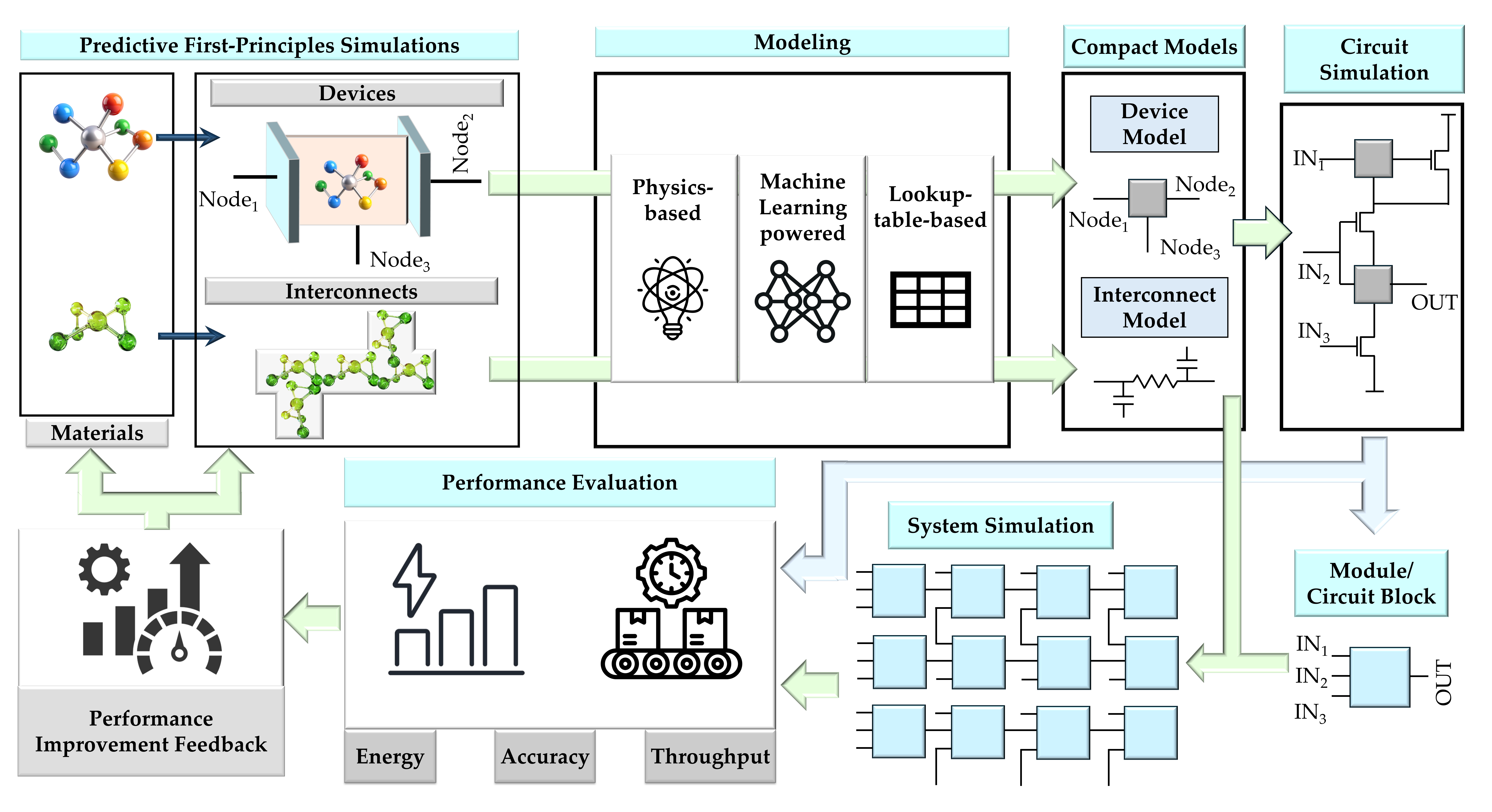}
    \caption{End-to-end multiscale co-design framework connecting first-principles physics to system-level performance. Predictive simulations provide material-specific electronic properties that feed into device- and interconnect-level modeling. These characteristics are translated into compact models using physics-based, machine-learning-powered, and lookup-table approaches, enabling accurate and SPICE-compatible representations of nanoscale behavior. The resulting device and interconnect models support circuit- and system-level simulations that generate workload-relevant metrics such as energy and throughput. These system-level outcomes supply feedback for iterative optimization of material selection, device geometries, and interconnect structures, establishing a predictive pathway for co-design across materials, devices, interconnects, circuits, and architectures. The transistor-level icons in the circuit-simulation block are simplified illustrative sketches and are not intended to represent complete complementary (CMOS) gate topologies.}
    \label{fig:framework}
\end{figure*}

The need for co-design across materials, devices, interconnects, circuit blocks, and system architectures can be addressed only through a consistent framework that connects physics-based simulations with performance metrics at higher levels of abstraction. Figure~\ref{fig:framework} summarizes the vision for such a comprehensive framework that can potentially enable multi-layer co-design. We emphasize that Figure~\ref{fig:framework} is a proposed research program, not a demonstrated pipeline: the stages demonstrated to date are the first-principles device and interconnect simulations of Sec.~\ref{sec:predictive} and, separately, system-level co-design studies based on compact device models\cite{Xiao22ontheaccuracy, Xiao22accurate_error_tolerant}; closing the full loop is precisely the program that this article advocates. It begins with predictive first-principles calculations that provide device characteristics including multi-terminal currents, transmission behavior, density of states, charge distributions, capacitances, mobility trends, and interfacial resistances. These quantities supply the fundamental inputs for modeling the active devices and the associated interconnect structures that determine energy and delay in large-scale computing fabrics.

To propagate the predicted device characteristics into circuit and system simulations, compact modeling becomes essential. Compact models convert quantum-transport behavior, interfacial phenomena, and interconnect parasitics into numerically efficient expressions that can be used in SPICE-compatible and higher-level simulators. Approaches based on physics-based formulations \cite{phymodel3,phymodel2,phymodel1}, machine-learning-assisted modeling \cite{mlmodel4,mlmodel3,mlmodel2,mlmodel1}, or lookup-table representations \cite{lutmodel3,lutmodel2,lutmodel1} allow complex nanoscale behavior, such as internal device states, non-ideal conduction regimes, and interconnect-dependent limitations, to be represented faithfully at the circuit level. With the emergence of increasingly complex materials and devices exhibiting rich, high-dimensional parameter spaces, machine-learning-based compact modeling has become especially attractive. After appropriate training using predictive first-principle simulations, ML models can function as digital twins of the underlying material or device, capturing intricate patterns in electronic behavior through optimization of the learning algorithm. 
With fast compact models, these predictive simulations can be used to project application-level solution quality (e.g, accuracy) ensuring that device-level effects do not compromise applications taking advantage of the proposed accelerators.
Beyond forward modeling, these ML-based representations also enable potential inverse design, where materials or device structures can be discovered or tuned to meet specific performance targets. Without such compact models, the information generated at the material and device scale cannot meaningfully influence the delay, throughput, and application accuracy metrics that define system performance.
Similarly, "device-agnostic" approaches to energy-efficiency of architectures cannot be meaningfully accurate, without the knowledge of device electrical characteristics, their variability and energy losses on leakage currents. It should not be forgotten that CMOS logic eliminates topological leakage (i.e. the direct $V_{DD}$–$V_{SS}$ path), but it cannot eliminate physical leakage (subthreshold, tunneling, junction, gate leakage\cite{GAAFET_TechRxiv}).

Interconnect modeling plays a particularly critical role in this workflow. In modern accelerators, a significant portion of the total energy is consumed in driving signals across local and global wires, and reductions in transistor energy alone do not automatically produce system-level gains if resistive or capacitive interconnect losses dominate. Novel conductors and low-dimensional materials may provide favorable resistivity or electrothermal properties at the nanoscale. However, unless these interconnect characteristics are incorporated into compact models and subsequently into circuit and system simulations, the resulting performance estimates will not reflect realistic operating conditions. Metrics such as energy per operation, signal delay, and achievable throughput depend jointly on both the device behavior and the interconnect behavior. Architectural estimators such as CACTI\cite{CACTI7} model cache, memory, and wiring energy and delay from technology-node-level analytical parameters; they consume, rather than produce, device- and interconnect-level inputs. The framework discussed here is complementary to such tools: predictive first-principles simulation supplies physically grounded parameters (effective resistivities, contact and interface resistances, capacitances, leakage components) for technologies for which empirical calibration data do not yet exist, and estimators of the CACTI class can then propagate these parameters to the architectural level.
Therefore, material- or device-level advantages must be evaluated within the full system context. 
\emph{High mobility materials, steep switching devices, or low contact/interface resistance become meaningful only when they translate into measurable improvements in circuit delay, fan-out capability, or workload-level energy efficiency.} 
Because every device ultimately operates within a network of interconnected components, nanoscale metrics alone are informative but not sufficient; system-level evaluation requires consistent propagation of material and device physics through the entire modeling hierarchy.
Moreover, recent simulation studies have shown that jointly optimizing for accuracy and efficiency on complex tasks requires device-to-algorithm co-design~\cite{Xiao22ontheaccuracy, Xiao22accurate_error_tolerant}.
Such system-level modeling requires fast compact models of device behavior which can be used millions-to-billions of times during a large scale application simulation. 
System-level simulations that employ compact models for both devices and interconnects can then provide workload-relevant metrics such as energy per matrix–vector multiplication,  throughput per watt, and task accuracy. These metrics supply the feedback needed for iterative optimization at the material and device stages. As illustrated in Figure~\ref{fig:framework}, this multi-stage workflow forms a direct and predictive pathway from first-principles physics to system-level performance and identifies the material choices, device geometries, doping profiles and interconnect structures that provide the most favorable gains for Beyond-Digital-CMOS accelerators.

\section{Conclusion}
The examples reviewed in this article establish, in concrete and quantitative terms, what predictive first-principles device simulations contribute to the co-design of energy-efficient AI hardware. For interconnects, open-boundary quantum transport reproduced the measured sheet resistances of Si:P $\delta$-layers across doping densities and thicknesses and revealed the quantized spatial structure of the current-carrying states (Figure~\ref{fig:Predictive_Simulaitons_Results_1}). For beyond-CMOS devices, the conductivity of $\delta$-layer tunnel junctions was predicted before the corresponding experiment and subsequently confirmed by it (Figure~\ref{fig:Predictive_Simulaitons_Results_2}a-b). For state-of-the-art CMOS, the same framework reproduced measured GAAFET $I$-$V$ characteristics and decomposed the gate-leakage components at realistic device dimensions (Figure~\ref{fig:Predictive_Simulaitons_Results_2}c-f), providing the device-level energy anchors used in Figure~\ref{fig:energy_ladder}. In all three cases the physical parameters were fixed independently of the device being predicted.

On this basis, predictive simulations supply three capabilities required for credible co-design of Beyond-Digital-CMOS accelerators. First, \emph{inverse design}, formulated in Sec.~\ref{sec:costs} as an optimization over the forward map of the simulator: determining geometry, materials, and doping profiles that meet circuit-supplied targets such as minimal energy per MAC. Second, \emph{multi-dimensional tradeoff analysis} that fitted models cannot provide outside their calibration range: the balance between leakage and drive strength, the constraints that confinement and tunneling place on feasible biasing and switching regimes, and the throughput limits set by interconnect $RC$. Third, \emph{faithful upward propagation of device physics} into circuit- and system-level studies via compact models, allowing material and device choices, under specific temperature and environmental conditions (e.g., radiation\cite{Google_space}), to be mapped onto MatMul energy, data-movement overhead, and ultimately token-level efficiency. Figure~\ref{fig:energy_ladder} frames the corresponding quantitative goal: the $1$--$100$~fJ-per-operation regime that neither improved digital switches nor device-agnostic architecture studies can reach or certify on their own.

Together, these three capabilities (inverse design, rigorous tradeoff analysis, and physics-consistent system extrapolation) define the distinctive role of predictive first-principles simulations within the co-design ecosystem.

\begin{acknowledgments}
This work is partially supported by the LDRD program at Sandia. The data that supports the findings of this study are available within the article. Sandia National Laboratories is a multimission laboratory managed and operated by National Technology and Engineering Solutions of Sandia, LLC., a wholly owned subsidiary of Honeywell International, Inc., for the U.S. Department of Energy’s National Nuclear Security Administration under contract DE-NA-0003525. This paper describes objective technical results and analysis. Any subjective views or opinions that might be expressed in the paper do not necessarily represent the views of the U.S. Department of Energy or the United States Government.
\end{acknowledgments}

\bibliography{main}

@book{AIbook,
  title     = {Artificial Intelligence: A Modern Approach},
  author    = {Stuart Russell and Peter Norvig},
  year      = {2021},
  edition   = {4th},
  publisher = {Pearson},
  address   = {Upper Saddle River, NJ},
  isbn      = {9780134610993}
}

@inproceedings{AI_power_consumption, series={FAccT ’24},
   title={Power Hungry Processing: Watts Driving the Cost of AI Deployment?},
   url={http://dx.doi.org/10.1145/3630106.3658542},
   DOI={10.1145/3630106.3658542},
   booktitle={The 2024 ACM Conference on Fairness, Accountability, and Transparency},
   publisher={ACM},
   author={Luccioni, Sasha and Jernite, Yacine and Strubell, Emma},
   year={2024},
   month=jun, pages={85–99},
   collection={FAccT ’24} }

@misc{BLOOM_power_consumption,
      title={Estimating the Carbon Footprint of BLOOM, a 176B Parameter Language Model}, 
      author={Alexandra Sasha Luccioni and Sylvain Viguier and Anne-Laure Ligozat},
      year={2022},
      eprint={2211.02001},
      archivePrefix={arXiv},
      primaryClass={cs.LG},
      url={https://arxiv.org/abs/2211.02001}, 
}

@article{2021LoihiSurvey,
  author    = {Mike Davies and Andreas Wild and Garrick Orchard and Yulia Sandamirskaya and Gabriel A. Fonseca Guerra and Prasad Joshi and Philipp Plank and Sumedh R. Risbud},
  title     = {Advancing Neuromorphic Computing With Loihi: A Survey of Results and Outlook},
  journal   = {Proceedings of the IEEE},
  volume    = {109},
  number    = {5},
  pages     = {911--934},
  year      = {2021},
  doi       = {10.1109/JPROC.2021.3067593},
  publisher = {IEEE}
}

@book{FNN,
  title     = {Deep Learning},
  author    = {Ian Goodfellow and Yoshua Bengio and Aaron Courville},
  year      = {2016},
  publisher = {MIT Press},
  address   = {Cambridge, MA},
  url       = {http://www.deeplearningbook.org}
}

@inproceedings{MegatronLM2021,
  title     = {Efficient Large-Scale Language Model Training on GPU Clusters Using Megatron-LM},
  author    = {Narayanan, Deepak and Shoeybi, Mohammad and Casper, Jared and LeGresley, Patrick and Patwary, Mostofa and Korthikanti, Vijay and Vainbrand, Dmitri and Kashinkunti, Prethvi and Bernauer, Julie and Catanzaro, Bryan and Phanishayee, Amar and others},
  booktitle = {Proceedings of the International Conference for High Performance Computing, Networking, Storage and Analysis (SC)},
  year      = {2021},
  eprint    = {2104.04473},
  archivePrefix = {arXiv},
  url       = {https://arxiv.org/abs/2104.04473}
}

@article{PaLM2022,
  title     = {PaLM: Scaling Language Modeling with Pathways},
  author    = {Chowdhery, Aakanksha and Narang, Sharan and Devlin, Jacob and Bosma, Maarten and Mishra, Gaurav and Roberts, Adam and Barham, Paul and Chung, Hyung Won and Sutton, Charles and Gehrmann, Sebastian and Schuh, Parker and others},
  journal   = {arXiv preprint arXiv:2204.02311},
  year      = {2022},
  url       = {https://arxiv.org/abs/2204.02311}
}

@article{Mamaluy2021_CommPhys,
title={Revealing quantum effects in highly conductive $\delta$-layer systems},
author={Mamaluy, Denis and Mendez, Juan P. and Gao, Xujiao and Misra, Shashank},
journal={Communications Physics},
year={2021},
month={Sep},
day={13},
volume={4},
number={1},
pages={205},
issn={2399-3650},
doi={10.1038/s42005-021-00705-1},
url={https://doi.org/10.1038/s42005-021-00705-1}
}

@article{DeltaLayer2022_SciRep,
  title     = {Conductivity and size quantization effects in semiconductor $\delta$-layer systems},
  author    = {Mendez, Juan P. and Mamaluy, Denis},
  journal   = {Scientific Reports},
  year      = {2022},
  doi       = {10.1038/s41598-022-20105-x},
  url       = {https://www.nature.com/articles/s41598-022-20105-x}
}

@article{TJ_experiment,
author = {Donnelly, Matthew B. and Munia, Mushita M. and Keizer, Joris G. and Chung, Yousun and Huq, A. M. Saffat-Ee and Osika, Edyta N. and Hsueh, Yu-Ling and Rahman, Rajib and Simmons, Michelle Y.},
title = {Multi-Scale Modeling of Tunneling in Nanoscale Atomically Precise Si:P Tunnel Junctions},
journal = {Advanced Functional Materials},
volume = {33},
number = {18},
pages = {2214011},
doi = {https://doi.org/10.1002/adfm.202214011},
url = {https://advanced.onlinelibrary.wiley.com/doi/abs/10.1002/adfm.202214011},
eprint = {https://advanced.onlinelibrary.wiley.com/doi/pdf/10.1002/adfm.202214011},
year = {2023}
}

@article{PhysRevApplied_2023,
  title = {Influence of imperfections on tunneling rate in $\ensuremath{\delta}$-layer junctions},
  author = {Mendez, Juan P. and Misra, Shashank and Mamaluy, Denis},
  journal = {Phys. Rev. Appl.},
  volume = {20},
  issue = {5},
  pages = {054021},
  numpages = {12},
  year = {2023},
  month = {Nov},
  publisher = {American Physical Society},
  doi = {10.1103/PhysRevApplied.20.054021},
  url = {https://link.aps.org/doi/10.1103/PhysRevApplied.20.054021}
}

@Article{Mendez2023,
author={Mendez, Juan P.
and Mamaluy, Denis},
title={Uncovering anisotropic effects of electric high-moment dipoles on the tunneling current in $\delta$-layer tunnel junctions},
journal={Scientific Reports},
year={2023},
month={Dec},
day={18},
volume={13},
number={1},
pages={22591},
issn={2045-2322},
doi={10.1038/s41598-023-49777-9},
url={https://doi.org/10.1038/s41598-023-49777-9}
}

@article{ACS_2025,
author = {Mendez, Juan P. and Mamaluy, Denis},
title = {Quantum Charge Sensing Using a Semiconductor Device Based on $\delta$-Layer Tunnel Junctions},
journal = {ACS Applied Electronic Materials},
volume = {7},
number = {11},
pages = {4898-4906},
year = {2025},
doi = {10.1021/acsaelm.5c00364}
}

@article{GAAFET_TechRxiv,
title={Gate-drain leakage enhanced by drain-induced dielectric barrier lowering in gate-all-around field-effect transistors},
journal={Journal of Applied Physics},
volume={139},
number={21},
pages={214501},
DOI={10.1063/5.0334284},
url={https://doi.org/10.1063/5.0334284},
publisher={AIP Publishing},
author={Mendez, Juan P. and Cariker, Coleman and Titze, Michael and Belianinov, Alex A. and Mamaluy, Denis},
year={2026},
month=jun }

@inproceedings{MamaluyMicDAT2024,
  title        = {Predictive Quantum Simulation and Device Physics of GAAFETs},
  author       = {Mamaluy, Denis and Mendez, Juan P. and Titze, Michael and Arghavani, Reza},
  booktitle    = {Proceedings of the 6th International Conference on Microelectronic Devices and Technologies (MicDAT '2024)},
  year         = {2024},
  url          = {https://sensorsportal.com/MicDAT_2024/MicDAT_2024_Proceedings.pdf}
}

@inproceedings{TPUv4,
  title     = {TPU v4: An Optically Reconfigurable Supercomputer for Machine Learning with Hardware Support for Embeddings},
  author    = {Norman P. Jouppi and George Kurian and Sheng Li and Peter Ma and Rahul Nagarajan and Lifeng Nai and Nishant Patil and Suvinay Subramanian and Andy Swing and Brian Towles and Cliff Young and Xiang Zhou and Zongwei Zhou and David Patterson},
  booktitle = {Proceedings of the 50th Annual International Symposium on Computer Architecture (ISCA)},
  year      = {2023},
  doi       = {10.1145/3579371.3589350},
  url       = {https://people.csail.mit.edu/suvinay/pubs/2023.tpu.isca.pdf}
}

@techreport{AMDmi325x,
  title     = {AMD Instinct MI325X Accelerator Datasheet},
  author    = {{Advanced Micro Devices, Inc.}},
  year      = {2024},
  institution = {AMD},
  url       = {https://www.amd.com/content/dam/amd/en/documents/instinct-tech-docs/product-briefs/instinct-mi325x-datasheet.pdf},
  note      = {Accessed August 2025}
}

@article{GPT,
  title={Improving Language Understanding by Generative Pre-Training},
  author={Radford, Alec and Narasimhan, Karthik and Salimans, Tim and Sutskever, Ilya},
  journal={OpenAI Blog},
  volume={1},
  number={8},
  year={2018},
  url={https://openai.com/research/language-unsupervised}
}

@inproceedings{GPT3,
  title={Language Models are Few-Shot Learners},
  author={Brown, Tom B. and Mann, Benjamin and Ryder, Nick and Subbiah, Melanie and Kaplan, Jared and Dhariwal, Prafulla and Neelakantan, Arvind and Shyam, Pranav and Sastry, Girish and Askell, Amanda and others},
  booktitle={Advances in Neural Information Processing Systems},
  volume={33},
  pages={1877--1901},
  year={2020},
  url={https://proceedings.neurips.cc/paper/2020/file/1457c0d6bfcb4967418bfb8ac142f64a-Paper.pdf}
}

@inproceedings{Multi-head-Attention,
  title={Attention Is All You Need},
  author={Vaswani, Ashish and Shazeer, Noam and Parmar, Niki and Uszkoreit, Jakob and Jones, Llion and Gomez, Aidan N and Kaiser, {\L}ukasz and Polosukhin, Illia},
  booktitle={Advances in Neural Information Processing Systems},
  volume={30},
  year={2017},
  url={https://papers.nips.cc/paper_files/paper/2017/file/3f5ee243547dee91fbd053c1c4a845aa-Paper.pdf}
}

@article{Mead1990Neuromorphic,
  author  = {Carver Mead},
  title   = {Neuromorphic Electronic Systems},
  journal = {Proceedings of the IEEE},
  volume  = {78},
  number  = {10},
  pages   = {1629--1636},
  year    = {1990},
  doi     = {10.1109/5.58356}
}

@article{Shannon1941Analog,
  author  = {Claude E. Shannon},
  title   = {Mathematical Theory of the Differential Analyzer},
  journal = {Journal of Mathematics and Physics},
  volume  = {20},
  number  = {1--4},
  pages   = {337--354},
  year    = {1941},
  doi     = {10.1002/sapm1941201337}
}

@article{Wave-parallel-computing,
  author    = {Yasushi Yuminaka and Yoshisato Sasaki and Takafumi Aoki and Tatsuo Higuchi},
  title     = {Design of Neural Networks Based on Wave-Parallel Computing Technique},
  journal   = {Analog Integrated Circuits and Signal Processing},
  volume    = {15},
  number    = {3},
  pages     = {315--327},
  year      = {1998},
  publisher = {Springer},
  doi       = {10.1023/A:1008274431412}
}

@article{mlmodel1,
author = {Hutchins, Jack and Alam, Shamiul and Rampini, Dana S. and Oripov, Bakhrom G. and McCaughan, Adam N. and Aziz, Ahmedullah},
doi = {10.1038/s41598-024-56779-8},
isbn = {0123456789},
issn = {2045-2322},
journal = {Scientific Reports 2024 14:1},
month = {mar},
number = {1},
pages = {1--9},
pmid = {38493250},
publisher = {Nature Publishing Group},
title = {{Machine learning-powered compact modeling of stochastic electronic devices using mixture density networks}},
url = {https://www.nature.com/articles/s41598-024-56779-8},
volume = {14},
year = {2024}
}

@article{phymodel1,
author = {Islam, Md Mazharul and Alam, Shamiul and Hossain, Md Shafayat and Aziz, Ahmedullah},
doi = {10.1109/ACCESS.2024.3363645},
issn = {21693536},
journal = {IEEE Access},
pages = {23200--23205},
publisher = {Institute of Electrical and Electronics Engineers Inc.},
title = {{Compact Model of a Topological Transistor}},
url = {https://ieeexplore.ieee.org/abstract/document/10423782},
volume = {12},
year = {2024}
}

@article{mlmodel2,
author = {Udoy, Md Rahatul Islam and Hutchins, Jack and Alam, Shamiul and Schuman, Catherine and Aziz, Ahmedullah},
doi = {10.1109/JXCDC.2025.3624662},
issn = {2329-9231},
journal = {IEEE Journal on Exploratory Solid-State Computational Devices and Circuits},
pages = {179--187},
title = {{Integrating Atomistic Insights With Circuit Simulations via Transformer-Driven Symbolic Regression}},
url = {https://ieeexplore.ieee.org/abstract/document/11215792},
volume = {11},
year = {2025}
}

@article{mlmodel3,
author = {Hutchins, Jack and Alam, Shamiul and Zeumault, Andre and Beckmann, Karsten and Cady, Nathaniel and Rose, Garrett S. and Aziz, Ahmedullah},
doi = {10.1109/ACCESS.2022.3218333},
issn = {21693536},
journal = {IEEE Access},
pages = {115513--115519},
publisher = {Institute of Electrical and Electronics Engineers Inc.},
title = {{A Generalized Workflow for Creating Machine Learning-Powered Compact Models for Multi-State Devices}},
volume = {10},
year = {2022}
}

@article{lutmodel1,
author = {Alam, Shamiul and Rampini, Dana S. and Oripov, Bakhrom G. and McCaughan, Adam N. and Aziz, Ahmedullah},
doi = {10.1063/5.0170187/2916063},
issn = {00036951},
journal = {Applied Physics Letters},
month = {oct},
number = {15},
publisher = {American Institute of Physics Inc.},
title = {{Cryogenic reconfigurable logic with superconducting heater cryotron: Enhancing area efficiency and enabling camouflaged processors}},
url = {/aip/apl/article/123/15/152603/2916063/Cryogenic-reconfigurable-logic-with},
volume = {123},
year = {2023}
}

@article{lutmodel2,
archivePrefix = {arXiv},
arxivId = {2408.01028},
author = {Islam, Md Mazharul and Alam, Shamiul and Udoy, Md Rahatul Islam and Hossain, Md Shafayat and Hamilton, Kathleen E and Aziz, Ahmedullah},
doi = {10.1063/5.0231749/3332040},
eprint = {2408.01028},
issn = {19319401},
journal = {Applied Physics Reviews},
month = {aug},
number = {1},
publisher = {AIP Publishing},
title = {{Harnessing Ferro-Valleytricity in Penta-Layer Rhombohedral Graphene for Memory and Compute}},
url = {http://arxiv.org/abs/2408.01028},
volume = {12},
year = {2024}
}

@article{phymodel2,
archivePrefix = {arXiv},
arxivId = {2203.16091},
author = {{Brosa Planella}, F. and Ai, W. and Boyce, A. M. and Ghosh, A. and Korotkin, I. and Sahu, S. and Sulzer, V. and Timms, R. and Tranter, T. G. and Zyskin, M. and Cooper, S. J. and Edge, J. S. and Foster, J. M. and Marinescu, M. and Wu, B. and Richardson, G.},
doi = {10.1088/2516-1083/AC7D31},
eprint = {2203.16091},
issn = {2516-1083},
journal = {Progress in Energy},
month = {jul},
number = {4},
pages = {042003},
publisher = {IOP Publishing},
title = {{A continuum of physics-based lithium-ion battery models reviewed}},
url = {https://iopscience.iop.org/article/10.1088/2516-1083/ac7d31 https://iopscience.iop.org/article/10.1088/2516-1083/ac7d31/meta},
volume = {4},
year = {2022}
}

@article{phymodel3,
author = {Sarker, Swapna and Kumar, Abhishek and Dasgupta, Avirup},
doi = {10.1016/j.sse.2025.109253},
issn = {00381101},
journal = {Solid-State Electronics},
month = {dec},
pages = {109253},
publisher = {Pergamon},
title = {{Physics-based compact model of subband energy for GAAFETs including corner rounding and geometric variability analysis utilizing Monte Carlo simulation}},
url = {http://dx.doi.org/10.1109/VTSA.2009.5159296},
volume = {230},
year = {2025}
}

@article{mlmodel4,
author = {Reuter, Maximilian and Wilm, Johannes and Kramer, Andreas and Bhattacharjee, Niladri and Beyer, Christoph and Trommer, Jens and Mikolajick, Thomas and Hofmann, Klaus},
doi = {10.1109/JEDS.2024.3386113},
issn = {21686734},
journal = {IEEE Journal of the Electron Devices Society},
pages = {310--317},
publisher = {Institute of Electrical and Electronics Engineers Inc.},
title = {{Machine Learning-Based Compact Model Design for Reconfigurable FETs}},
url = {https://ieeexplore.ieee.org/abstract/document/10494540},
volume = {12},
year = {2024}
}

@article{lutmodel3,
archivePrefix = {arXiv},
arxivId = {2301.01689},
author = {James, Aneek and Rizzo, Anthony and Wang, Yuyang and Novick, Asher and Wang, Songli and Parsons, Robert and Jang, Kaylx and Hattink, Maarten and Bergman, Keren},
doi = {10.1109/JLT.2023.3238847},
eprint = {2301.01689},
issn = {15582213},
journal = {Journal of Lightwave Technology},
month = {may},
number = {9},
pages = {2801--2814},
publisher = {Institute of Electrical and Electronics Engineers Inc.},
title = {{Process Variation-Aware Compact Model of Strip Waveguides for Photonic Circuit Simulation}},
url = {https://ieeexplore.ieee.org/abstract/document/10024277},
volume = {41},
year = {2023}
}

@misc{IEA2025_EnergyAI,
  author       = {{International Energy Agency}},
  title        = {Energy and AI},
  year         = {2025},
  publisher    = {IEA},
  address      = {Paris},
  howpublished = {\url{https://www.iea.org/reports/energy-and-ai}},
  note         = {Licence: CC BY 4.0}
}

@techreport{DOE_EES2_2024,
  title        = {Microelectronics Energy Efficiency Scaling for 2 Decades (EES2) RD\&D Roadmap},
  institution  = {U.S. Department of Energy, Advanced Materials and Manufacturing Technologies Office (AMMTO)},
  year         = {2024},
  url          = {https://www.energy.gov/sites/default/files/2024-08/Draft_EES2_Roadmap_AMMTO_August29_2024.pdf},
  note         = {Draft report}
}

@techreport{DOE_AI_DataCenter_2024,
  title        = {Recommendations on Powering Artificial Intelligence and Data Center Infrastructure},
  institution  = {U.S. Department of Energy},
  year         = {2024},
  month        = {September},
  url          = {https://www.energy.gov/},
  note         = {DOE report outlining energy demand impacts of AI-driven data centers},
}

@article{ThermalPerformanceDataCenters,
  title   = {Improving Thermal Performance in Data Centers Based on Numerical Simulations},
  author  = {Guo, Yinjie and Zhao, Chunyu and Gao, Hao and Shen, Cheng and Fu, Xu},
  journal = {Buildings},
  volume  = {14},
  number  = {5},
  pages   = {1416},
  year    = {2024},
  doi     = {10.3390/buildings14051416},
  note    = {Discusses the increasing prominence of heat dissipation issues in data centers driven by cloud computing and AI workloads},
}

@inproceedings{Interconnects_power,
  title={Understanding the Power Consumption of On-Chip Interconnects: Experimental Characterization, Modeling, and Analysis on Real Hardware},
  author={Adhinarayanan, Vignesh and Pauly, Indrani and Greathouse, Joseph L. and Huang, Wei and Pattnaik, Ashutosh and Feng, Wu-chun},
  booktitle={2016 IEEE International Symposium on Workload Characterization (IISWC)},
  year={2016},
  pages={1--10},
  doi={10.1109/IISWC.2016.7581271}
}

@Article{Wang2021,
author={Wang, Yin
and Tang, Hongwei
and Xie, Yufeng
and Chen, Xinyu
and Ma, Shunli
and Sun, Zhengzong
and Sun, Qingqing
and Chen, Lin
and Zhu, Hao
and Wan, Jing
and Xu, Zihan
and Zhang, David Wei
and Zhou, Peng
and Bao, Wenzhong},
title={An in-memory computing architecture based on two-dimensional semiconductors for multiply-accumulate operations},
journal={Nature Communications},
year={2021},
month={Jun},
day={07},
volume={12},
number={1},
pages={3347},
issn={2041-1723},
doi={10.1038/s41467-021-23719-3},
url={https://doi.org/10.1038/s41467-021-23719-3}
}

@article{Ripa2021_EnergyMetabolismPostIndustrial,
  title   = {The energy metabolism of post-industrial economies: A framework to account for externalization across scales},
  author  = {Ripa, M. and Di Felice, L. J. and Giampietro, M.},
  journal = {Energy},
  volume  = {236},
  pages   = {121–943},
  year    = {2021},
  url     = {https://zenodo.org/records/4069166},
  note    = {Analyzes structural limits of energy production in post-industrial societies and their reliance on externalized energy flows},
}

@article{Co-design1,
  author    = {Wayne Wolf},
  title     = {Hardware--Software Co-Design of Embedded Systems},
  journal   = {Proceedings of the IEEE},
  volume    = {82},
  number    = {7},
  pages     = {967--989},
  year      = {1994},
  doi       = {10.1109/5.293155}
}

@article{Co-design2,
  author  = {Hill, Mark D. and Marty, Michael R.},
  title   = {Amdahl's Law in the Multicore Era},
  journal = {IEEE Computer},
  volume  = {41},
  number  = {7},
  pages   = {33--38},
  year    = {2008},
  doi     = {10.1109/MC.2008.209}
}

@inproceedings{WanlassSah1963,
  author    = {Wanlass, Frank M. and Sah, Chih{-}Tang},
  title     = {Nanowatt Logic Using Field-Effect Metal-Oxide Semiconductor Triodes},
  booktitle = {Proceedings of the IEEE International Solid-State Circuits Conference (ISSCC)},
  year      = {1963},
  pages     = {32--33},
  address   = {Philadelphia, PA},
  note      = {ISSCC Digest of Technical Papers}
}

@article{Keldysh:1965,
author = {Keldysh, L. V.},
title = {Diagram Technique for Nonequilibrium Processes},
journal = {Sov. Phys. J. Exp. Theor. Phys.},
volume = {20},
number = {4},
pages = {1018},
year = {1965}
}

@book{Datta:1997,
  title={Electronic transport in mesoscopic systems},
  author={Datta, Supriyo},
  year={1997},
  publisher={Cambridge university press}
}

@article{Mamaluy:2003,
author = {Mamaluy,D.  and Sabathil,M.  and Vogl,P. },
title = {Efficient method for the calculation of ballistic quantum transport},
journal = {J. Appl. Phys.},
volume = {93},
number = {8},
pages = {4628-4633},
year = {2003},
doi = {10.1063/1.1560567},
URL = { https://doi.org/10.1063/1.1560567},
eprint = {https://doi.org/10.1063/1.1560567}
}

@article{Mamaluy_2004,
	doi = {10.1088/0268-1242/19/4/042},
	url = {https://doi.org/10.1088/0268-1242/19/4/042},
	year = 2004,
	month = {mar},
	publisher = {{IOP} Publishing},
	volume = {19},
	number = {4},
	pages = {S118--S121},
	author = {Denis Mamaluy and Anand Mannargudi and Dragica Vasileska and Matthias Sabathil and Peter Vogl},
	title = {Contact block reduction method and its application to a 10 nm {MOSFET} device},
	journal = {Semiconductor Science and Technology}
}

@article{Sabathil_2004,
	doi = {10.1088/0268-1242/19/4/048},
	url = {https://doi.org/10.1088/0268-1242/19/4/048},
	year = 2004,
	month = {mar},
	publisher = {{IOP} Publishing},
	volume = {19},
	number = {4},
	pages = {S137--S138},
	author = {M Sabathil and D Mamaluy and P Vogl},
	title = {Prediction of a realistic quantum logic gate using the contact block reduction method},
	journal = {Semiconductor Science and Technology},
}

@article{Mamaluy:2005,
  title = {Contact block reduction method for ballistic transport and carrier densities of open nanostructures},
  author = {Mamaluy, D. and Vasileska, D. and Sabathil, M. and Zibold, T. and Vogl, P.},
  journal = {Phys. Rev. B},
  volume = {71},
  issue = {24},
  pages = {245321},
  numpages = {14},
  year = {2005},
  month = {Jun},
  publisher = {American Physical Society},
  doi = {10.1103/PhysRevB.71.245321},
  url = {https://link.aps.org/doi/10.1103/PhysRevB.71.245321}
}

@ARTICLE{Khan:2007,
author={H. R. {Khan} and D. {Mamaluy} and D. {Vasileska}},
journal={IEEE T. Electron Dev.},
title={Quantum Transport Simulation of Experimentally Fabricated Nano-FinFET},
year={2007},
volume={54},
number={4},
pages={784-796},
doi={10.1109/TED.2007.892353},
ISSN={1557-9646},
month={April}
}

@article{Khan_2008,
	doi = {10.1088/1742-6596/107/1/012007},
	url = {https://doi.org/10.1088/1742-6596/107/1/012007},
	year = 2008,
	month = {mar},
	publisher = {{IOP} Publishing},
	volume = {107},
	pages = {012007},
	author = {H R Khan and D Mamaluy and D Vasileska},
	title = {3D {NEGF} quantum transport simulator for modeling ballistic transport in nano {FinFETs}},
	journal = {Journal of Physics: Conference Series}
}

@article{Gao:2014,
author = {Gao,X.  and Mamaluy,D.  and Nielsen,E.  and Young,R. W.  and Shirkhorshidian,A.  and Lilly,M. P.  and Bishop,N. C.  and Carroll,M. S.  and Muller,R. P. },
title = {Efficient self-consistent quantum transport simulator for quantum devices},
journal = {J. Appl. Phys.},
volume = {115},
number = {13},
pages = {133707},
year = {2014},
doi = {10.1063/1.4870288},
URL = { https://doi.org/10.1063/1.4870288},
}

@INPROCEEDINGS{Mendez:2021,
  author={Mendez, Juan P. and Mamaluy, Denis and Gao, Xujiao and Misra, Shashank},
  booktitle={2021 International Conference on Simulation of Semiconductor Processes and Devices (SISPAD)}, 
  title={Quantum Transport Simulations for Si:P $\delta$-layer Tunnel Junctions}, 
  year={2021},
  volume={},
  number={},
  pages={210-214},
  doi={10.1109/SISPAD54002.2021.9592565},
  url={https://doi.org/10.1109/SISPAD54002.2021.9592565}
  }

@misc{Google_space,
      title={Towards a future space-based, highly scalable AI infrastructure system design}, 
      author={Blaise Agüera y Arcas and Travis Beals and Maria Biggs and Jessica V. Bloom and Thomas Fischbacher and Konstantin Gromov and Urs Köster and Rishiraj Pravahan and James Manyika},
      year={2025},
      eprint={2511.19468},
      archivePrefix={arXiv},
      url={https://arxiv.org/abs/2511.19468}, 
}

@ARTICLE{Xiao22ontheaccuracy,
  author={Xiao, T. Patrick and Feinberg, Ben and Bennett, Christopher H. and Prabhakar, Venkatraman and Saxena, Prashant and Agrawal, Vineet and Agarwal, Sapan and Marinella, Matthew J.},
  journal={IEEE Circuits and Systems Magazine}, 
  title={On the Accuracy of Analog Neural Network Inference Accelerators}, 
  year={2022},
  volume={22},
  number={4},
  pages={26-48},
  doi={10.1109/MCAS.2022.3214409},
  ISSN={1558-0830},
  month={Fourthquarter},
}

@ARTICLE{Xiao22accurate_error_tolerant,
  author={Xiao, T. Patrick and Feinberg, Ben and Bennett, Christopher H. and Agrawal, Vineet and Saxena, Prashant and Prabhakar, Venkatraman and Ramkumar, Krishnaswamy and Medu, Harsha and Raghavan, Vijay and Chettuvetty, Ramesh and Agarwal, Sapan and Marinella, Matthew J.},
  journal={IEEE Transactions on Circuits and Systems I: Regular Papers}, 
  title={An Accurate, Error-Tolerant, and Energy-Efficient Neural Network Inference Engine Based on SONOS Analog Memory}, 
  year={2022},
  volume={69},
  number={4},
  pages={1480-1493},
  doi={10.1109/TCSI.2021.3134313}}

@article{Quantum_dominates_1,
  author    = {G. Milano and M. Aono and L. Boarino and U. Celano and T. Hasegawa and M. Kozicki and S. Majumdar and M. Menghini and E. Miranda and C. Ricciardi and S. Tappertzhofen},
  title     = {Quantum Conductance in Memristive Devices: Fundamentals, Developments, and Applications},
  journal   = {Advanced Materials},
  volume    = {34},
  number    = {32},
  pages     = {e2201248},
  year      = {2022},
  doi       = {10.1002/adma.202201248},
  note      = {Review on quantum conductance dominating transport in memristive nanodevices.},
}

@article{Quantum_dominates_2,
  author    = {Ghada Badawy and E. P. A. M. Bakkers},
  title     = {Electronic Transport and Quantum Phenomena in Nanowires},
  journal   = {Chemical Reviews},
  volume    = {124},
  number    = {5},
  pages     = {2419--2440},
  year      = {2024},
  doi       = {10.1021/acs.chemrev.3c00656},
  note      = {Comprehensive review of quantum transport phenomena in nanowires, where quantum effects dominate conduction.},
}

@article{Quantum_dominates_3,
  author    = {K. Y. Kim and H.-H. Park and S. Jin and U. Kwon and W. Choi and D. S. Kim},
  title     = {Quantum transport through a constriction in nanosheet gate‐all‐around transistors},
  journal   = {Communications Engineering},
  volume    = {4},
  article   = {92},
  year      = {2025},
  doi       = {10.1038/s44172-025-00435-0},
  note      = {Shows quantum mechanical tunnelling and confinement effects essential in conduction of scaled transistors.},
}

@article{Goh:2006,
  title = {Influence of doping density on electronic transport in degenerate Si:P $\ensuremath{\delta}$-doped layers},
  author = {Goh, K. E. J. and Oberbeck, L. and Simmons, M. Y. and Hamilton, A. R. and Butcher, M. J.},
  journal = {Phys. Rev. B},
  volume = {73},
  issue = {3},
  pages = {035401},
  numpages = {6},
  year = {2006},
  month = {Jan},
  publisher = {American Physical Society},
  doi = {10.1103/PhysRevB.73.035401},
  url = {https://link.aps.org/doi/10.1103/PhysRevB.73.035401}
}

@article{Goh:2009,
author = {Goh,K. E. J.  and Simmons,M. Y. },
title = {Impact of Si growth rate on coherent electron transport in Si:P delta-doped devices},
journal = {Appl. Phys. Lett.},
volume = {95},
number = {14},
pages = {142104},
year = {2009},
doi = {10.1063/1.3245313},
URL = { https://doi.org/10.1063/1.3245313},
eprint = { https://doi.org/10.1063/1.3245313}
}

@article{Reusch:2008,
author = {Reusch,T. C. G.  and Goh,K. E. J.  and Pok,W.  and Lo,W.-C. N.  and McKibbin,S. R.  and Simmons,M. Y. },
title = {Morphology and electrical conduction of Si:P $\delta$-doped layers on vicinal Si(001)},
journal = {J. Appl. Phys.},
volume = {104},
number = {6},
pages = {066104},
year = {2008},
doi = {10.1063/1.2977750},
URL = { https://doi.org/10.1063/1.2977750},
eprint = { https://doi.org/10.1063/1.2977750}
}

@article{McKibbin:2014,
author = {McKibbin,S. R.  and Polley\textbf{},C. M.  and Scappucci,G.  and Keizer,J. G.  and Simmons,M. Y. },
title = {Low resistivity, super-saturation phosphorus-in-silicon monolayer doping},
journal = {Appl. Phys. Lett.},
volume = {104},
number = {12},
pages = {123502},
year = {2014},
doi = {10.1063/1.4869111},
URL = { https://doi.org/10.1063/1.4869111},
eprint = { https://doi.org/10.1063/1.4869111}
}

@INPROCEEDINGS{Huang:2008,
  author={Huang, Qiaojian and Lilley, Carmen M. and Bode, Matthias and Divan, Ralu S.},
  booktitle={2008 8th IEEE Conference on Nanotechnology}, 
  title={Electrical Properties of Cu Nanowires}, 
  year={2008},
  volume={},
  number={},
  pages={549-552},
  doi={10.1109/NANO.2008.163}}

@article{Steinhogl:2002,
  title = {Size-dependent resistivity of metallic wires in the mesoscopic range},
  author = {Steinh\"ogl, Werner and Schindler, G\"unther and Steinlesberger, Gernot and Engelhardt, Manfred},
  journal = {Phys. Rev. B},
  volume = {66},
  issue = {7},
  pages = {075414},
  numpages = {4},
  year = {2002},
  month = {Aug},
  publisher = {American Physical Society},
  doi = {10.1103/PhysRevB.66.075414},
  url = {https://link.aps.org/doi/10.1103/PhysRevB.66.075414}
}

@article{Josell:2009,
   author = "Josell, Daniel and Brongersma, Sywert H. and Tőkei, Zsolt",
   title = "Size-Dependent Resistivity in Nanoscale Interconnects", 
   journal= "Annual Review of Materials Research",
   year = "2009",
   volume = "39",
   number = "Volume 39, 2009",
   pages = "231-254",
   doi = "https://doi.org/10.1146/annurev-matsci-082908-145415",
   url = "https://www.annualreviews.org/content/journals/10.1146/annurev-matsci-082908-145415",
   publisher = "Annual Reviews",
   issn = "1545-4118",
   type = "Journal Article"
  }

@article{Graham:2022,
    author = {Graham, R. L. and Alers, G. B. and Mountsier, T. and Shamma, N. and Dhuey, S. and Cabrini, S. and Geiss, R. H. and Read, D. T. and Peddeti, S.},
    title = {Resistivity dominated by surface scattering in sub-50 nm Cu wires},
    journal = {Applied Physics Letters},
    volume = {96},
    number = {4},
    pages = {042116},
    year = {2010},
    month = {01},
    issn = {0003-6951},
    doi = {10.1063/1.3292022},
    url = {https://doi.org/10.1063/1.3292022}
}

@ARTICLE{Dutta:2018,
  author={Dutta, Shibesh and Beyne, Sofie and Gupta, Anshul and Kundu, Shreya and Van Elshocht, Sven and Bender, Hugo and Jamieson, Geraldine and Vandervorst, Wilfried and B\"ommels, J\"urgen and Wilson, Christopher J. and T\"okei, Zsolt and Adelmann, Christoph},
  journal={IEEE Electron Device Letters}, 
  title={Sub-100 nm2 Cobalt Interconnects}, 
  year={2018},
  volume={39},
  number={5},
  pages={731-734},
  doi={10.1109/LED.2018.2821923}}

@article{Hu:2023,
  title = {Size effect of resistivity due to surface roughness scattering in alternative interconnect metals: Cu, Co, Ru, and Mo},
  author = {Hu, Chaoyu and Zhang, Yu and Chen, Zhiyi and Zhang, Qingyun and Zhu, Jianjun and Hu, Shaojian and Ke, Youqi},
  journal = {Phys. Rev. B},
  volume = {107},
  issue = {19},
  pages = {195422},
  numpages = {11},
  year = {2023},
  month = {May},
  publisher = {American Physical Society},
  doi = {10.1103/PhysRevB.107.195422},
  url = {https://link.aps.org/doi/10.1103/PhysRevB.107.195422}
}

@inproceedings{Horowitz2014,
author={Horowitz, Mark},
title={1.1 Computing's energy problem (and what we can do about it)},
booktitle={2014 IEEE International Solid-State Circuits Conference Digest of Technical Papers (ISSCC)},
pages={10--14},
year={2014},
DOI={10.1109/ISSCC.2014.6757323} }

@article{CACTI7,
author={Balasubramonian, Rajeev and Kahng, Andrew B. and Muralimanohar, Naveen and Shafiee, Ali and Srinivas, Vaishnav},
title={{CACTI 7}: New tools for interconnect exploration in innovative off-chip memories},
journal={ACM Transactions on Architecture and Code Optimization},
volume={14},
number={2},
pages={14},
year={2017},
DOI={10.1145/3085572} }

@article{Luisier2006,
author={Luisier, Mathieu and Schenk, Andreas and Fichtner, Wolfgang and Klimeck, Gerhard},
title={Atomistic simulation of nanowires in the $sp^3d^5s^*$ tight-binding formalism: From boundary conditions to strain calculations},
journal={Phys. Rev. B},
volume={74},
pages={205323},
year={2006},
DOI={10.1103/PhysRevB.74.205323} }

@article{Steiger2011,
author={Steiger, Sebastian and Povolotskyi, Michael and Park, Hong-Hyun and Kubis, Tillmann and Klimeck, Gerhard},
title={{NEMO5}: A parallel multiscale nanoelectronics modeling tool},
journal={IEEE Transactions on Nanotechnology},
volume={10},
number={6},
pages={1464--1474},
year={2011},
DOI={10.1109/TNANO.2011.2166164} }

@article{Brandbyge2002,
author={Brandbyge, Mads and Mozos, Jos\'e-Luis and Ordej\'on, Pablo and Taylor, Jeremy and Stokbro, Kurt},
title={Density-functional method for nonequilibrium electron transport},
journal={Phys. Rev. B},
volume={65},
pages={165401},
year={2002},
DOI={10.1103/PhysRevB.65.165401} }

@article{Papior2017,
author={Papior, Nick and Lorente, Nicol\'as and Frederiksen, Thomas and Garc\'ia, Alberto and Brandbyge, Mads},
title={Improvements on non-equilibrium and transport Green function techniques: The next-generation transiesta},
journal={Computer Physics Communications},
volume={212},
pages={8--24},
year={2017},
DOI={10.1016/j.cpc.2016.09.022} }

@article{Smidstrup2020,
author={Smidstrup, S{\o}ren and Markussen, Troels and Vancraeyveld, Pieter and Wellendorff, Jess and Schneider, Julian and Gunst, Tue and Verstichel, Brecht and Stradi, Daniele and Khomyakov, Petr A. and Vej-Hansen, Ulrik G. and others},
title={{QuantumATK}: an integrated platform of electronic and atomic-scale modelling tools},
journal={Journal of Physics: Condensed Matter},
volume={32},
pages={015901},
year={2020},
DOI={10.1088/1361-648X/ab4007} }

@article{MayadasShatzkes1970,
author={Mayadas, A. F. and Shatzkes, M.},
title={Electrical-resistivity model for polycrystalline films: the case of arbitrary reflection at external surfaces},
journal={Phys. Rev. B},
volume={1},
pages={1382--1389},
year={1970},
DOI={10.1103/PhysRevB.1.1382} }

@article{FET_downscaling_limit,
author={Mamaluy, Denis and Gao, Xujiao},
title={The fundamental downscaling limit of field effect transistors},
journal={Appl. Phys. Lett.},
volume={106},
pages={193503},
year={2015},
DOI={10.1063/1.4919871} }

\end{document}